# SOVEREIGN WEALTH FUNDS: MAIN ACTIVITY TRENDS

**Oksana Mamina,**
*Associate Professor, Russian University of Transport,*
*Moscow, Russia*

**Alexander Barannikov,**
*Associate Professor, Russian Academy of National Economy and Public Administration,*
*Moscow, Russia*

**Ludmila Gruzdeva,**
*Associate Professor, Russian University of Transport,*
*Moscow, Russia*

**Abstract**

Sovereign wealth funds are created in those countries whose budget is highly dependent on market factors, usually world commodity prices. At the same time, these funds are large institutional investors. An analysis of the nature of investments by the State Pension Fund Global of Norway showed that investments of the Fund are based on a seven-level model of diversifying its investments. This model can also be applied to the investments of the National Wealth Fund of Russia to increase its profitability.

## 1. Introduction

The dependence of a number of countries around the world on price fluctuations in global commodity markets has led to the need to create reserves or funds. Many countries in the world founded sovereign wealth funds. These funds are used to maintain financial stability in times of crisis and as a tool for global investment. The main task of the sovereign wealth fund is to cover the budget deficit in the case of the adverse economic conditions or the accumulation of excess export earnings. These revenues can be invested in various projects, for example, in the construction of infrastructure, etc.

## 2. Sovereign wealth funds around the world

Since the creation of the first two sovereign wealth funds (Kuwaiti Investment Council, 1953, and the Kiribati Reserve Revenue Balancing Fund, 1965), their number has increased significantly to about 70 funds. Sovereign wealth funds are established and operated in many countries around the world. As of June 2018 total assets of the funds amounted to 7820 million dollars. [1]. Revenues from the export of commodities, mainly oil and oil products, are the basis for the formation of the assets of most of the funds in the world. In terms of assets, the three largest sovereign wealth funds include the funds of Norway, China and the United Arab Emirates (Abu Dhabi). The National Welfare Fund of the Russian Federation ranks 18th in the world (see Table 1) [1].

The objectives of the National Welfare Fund are to ensure the co-financing of voluntary pension savings of citizens of the Russian Federation and to ensure balance (deficit coverage) of the budget of the Pension Fund of the Russian Federation [2]. As of October 1, 2018 its assets amount to 5004.49 billion rubles (USD 76.3) or 5.1% of GDP. Linabourg-Maduell Index (Index in table 1) shows the transparency level.

*Table 1. Assets of the largest sovereign wealth funds in 2018 (bln. $)*

| Country | Name of the fund | Assets | Year | Base | Index |
|---|---|---|---|---|---|
| 1. Norway | Government Pension Fund-Global | 1058.05 | 1990 | oil | 10 |
| 2. China | China Investment Corporation | 941.4 | 2007 | noncommodity | 8 |
| 3. UAE, Abu-Dhabi | Abu Dhabi Investment Authority | 683 | 1976 | oil | 6 |
| 4. Kuwait | Kuwait Investment Authority | 592 | 1953 | oil | 6 |
| 5. China Hong Kong | Hong Kong Monetary Authority Investment Portfolio | 522.6 | 1993 | noncommodity | 8 |
| 6. Saudi Arabia | SAMA Foreign Holdings | 515.6 | 1952 | oil | 4 |
| 7. China | SAFE Investment Company | 441 | 1997 | noncommodity | 4 |
| 8. Singapore | Government of Singapore Investment Corporation | 390 | 1981 | noncommodity | 6 |
| 9. Singapore | Temasek Holdings | 375 | 1974 | noncommodity | 10 |
| 10. Saudi Arabia | Public Invest-ment Fund | 360 | 2008 | oil | 5 |
| 11. Qatar | Qatar Investment Authority | 320 | 2005 | oil and gas | 5 |
| 12. China | National Social Security Fund | 295 | 2000 | noncommodity | 5 |
| 13. UAE Dubai | Investment Corporation of Dubai | 229.8 | 2006 | noncommodity | 5 |
| 14. UAE Abu-Dhabi | Mudabala Investment Company | 226 | 2002 | oil | 10 |
| 15. South Korea | Korea Investment Corporation | 134.1 | 2005 | noncommodity | 9 |

| 16. Australia | Australian Future Fund | 107.7 | 2006 | noncom-modity | 10 |
| 17. Iran | National Development Fund of Iran | 91 | 2011 | oil and gas | 5 |
| 18. Russia | National Wealth Fund | 77.2 | 2008 | oil | 5 |

*Source: [1].*

Sovereign funds have significant differences in their investment strategy and the degree of transparency and disclosure. Thus, the foundations of Norway, Singapore, the United Arab Emirates and Australia have the highest index of transparency Linaburg-Maduell, the National Welfare Fund of the Russian Federation has an average value of this index of 5 (see Table 1). According to one of the leaders of Global Insight, Jan Randolph, sovereign funds are the newest leading force in the global financial market, which is replacing hedge funds and private investment funds [3].

### 3. Government pension fund of Norway

The Government Pension Fund-Global of Norway was one of the most successful sovereign funds. 66.6 percent of the fund's assets were invested in the shares, 30.8 percent accounted for investments with fixed income (mainly bonds), 2.6 percent - for real estate. About 40 percent of the fund's funds are accounted for by government funding, the rest was obtained through successful investments.

In 2017, it earned more than one trillion kronor (US $ 131.5 billion at the current exchange rate) from investments. The yield on the fund's investments was 13.7 percent, according to data from the annual report published on February 27. On average for all time investments since 1998, the yield was 6.1 percent. In 2016, the yield was 7 percent [4].

The most successful were investments in stocks. The yield on these securities amounted to 19.4 percent in 2017. The most profitable were shares of technology companies. On them, the fund earned 32.4 percent of income from invested funds. Investments in stocks of companies producing basic materials showed a yield of 27.1 percent.

Investments in metallurgy and chemical industry turned out to be the most profitable. The rise in prices for metals and chemical compounds was due to steady demand, combined with a reduction in production capacity in China.

A good result also brought investment in real estate - 7.5 percent yield. Bonds, as a less risky instrument, showed a low yield of 3.3 percent. The capitalization of the fund by the end of 2017 amounted to 8.488 trillion kronor (1.085 trillion dollars).

The Governmental Pension Fund Global Norway has the following seven levels of diversification of investments of financial resources: 1) stocks and bonds; 2) regions (Europe, America, Asia / Oceania); 3) currencies; 4) countries; 5) industries; 6) companies / organizations; 7) corporate / non-corporate issuers.

- the first level is investment in stocks and bonds. Preference is given to bonds (about 60% of the portfolio) as the most stable and reliable securities. The fund adheres to the statutory ratios between stocks (30–50%) and bonds (70–50%).

- the second level represents investments by region (Europe, America, Asia and Oceania). Here the advantage is given to Europe as the most dynamically developing region, which is in close proximity to Norway. Legislatively are established the limits on the investment of shares and bonds by region.

- the third level is investment in currencies. The fund invests in various currencies, and then conducts final settlements in Norwegian kronas. Recalculation into a single currency is very important, since the profitability of the GPF differs, when it is expressed in different currencies.

- the fourth level is the investment by country, with the priority belongs to the countries of Europe. For investments in stocks and bonds, various lists of countries where the financial resources of the fund are invested are separately established.

- the fifth level is investment by industry. It has the advantage of financial and banking sector, energy enterprises, consumer goods, telecommunications and communications.

- the sixth level is the investment in securities of companies-issuers. It is accepted that the amount of investment in each individual company may not exceed 3% of its voting shares or share capital.

- the seventh level is the investment by type of issuer of securities (corporate / unincorporated securities). It is present only in investments in bonds, since all investments in stocks are related to corporate securities.

The revenues of the fund portfolio are calculated according to market prices. The State Bank of Norway checks the market value of a portfolio and calculates weighted average return only at the end of the month. For this, the method of weighted average monetary valuation (modified Ditz method) is used, which is determined using the special formula:

$$R_M = \left[\left(\frac{MV_E - \frac{\sum_i i \cdot K_i}{T}}{MV_B + \frac{\sum_i (T-i) \cdot K_i}{T}}\right) - 1\right] \cdot 100 \quad (1)$$

where RM is the weighted average cash income for the period,%;

MVB and MVE - portfolio value, respectively, at the beginning and at the end of the period;

T - numbers of days of the period;
i - numbers of days attributable to cash flow Ki;
Ki is the value of the cash flow on the i-th day.

### 4. National Wealth Fund of Russia

The National Wealth Fund is part of the federal budget of the Russian Federation. The fund is intended to be part of a sustainable mechanism for providing pensions to citizens of the Russian Federation for the long term. The objectives of the National Welfare Fund are to ensure the co-financing of voluntary pension savings of citizens of the Russian Federation and to ensure balance (deficit coverage) of the budget of the Pension Fund of the Russian Federation [2].

From January to November 2018, the National Wealth Fund increased by 1,219 billion rubles. The fund grew 5 months out of 11 and reached its maximum size in a year in September, when it rose to 3,699 billion rubles. The dynamics of the Fund is presented in Table 2.

*Table 2. Volume of the National Wealth Fund*

| Date | in Billion US dollars | in Billion rubles | as Percentages of GDP |
|---|---|---|---|
| 01.11.2018 | 75,60 | 4 972,45 | 5,1% |
| 01.10.2018 | 76,30 | 5 004,49 | 5,1% |
| 01.09.2018 | 75,79 | 5 160,28 | 5,3% |
| 01.08.2018 | 77,16 | 4 844,38 | 4,9% |
| 01.07.2018 | 77,11 | 4 839,26 | 4,9% |
| 01.06.2018 | 62,75 | 3 927,58 | 4,0% |
| 01.05.2018 | 63,91 | 3 962,67 | 4,0% |
| 01.04.2018 | 65,88 | 3 772,83 | 3,8% |
| 01.03.2018 | 66,44 | 3 698,96 | 3,8% |
| 01.02.2018 | 66,26 | 3 729,71 | 3,8% |
| 01.01.2018 | 65,15 | 3 752,94 | 3,8% |
| 01.12.2017 | 66,94 | 3 904,76 | 4,2% |
| 01.11.2017 | 69,36 | 4 013,81 | 4,4% |
| 01.10.2017 | 72,57 | 4 210,36 | 4,6% |
| 01.09.2017 | 75,36 | 4 425,68 | 4,8% |
| 01.08.2017 | 74,72 | 4 449,35 | 4,8% |
| 01.07.2017 | 74,22 | 4 385,49 | 4,8% |

*Source: [2].*

The assets of the Fund can be invested in following ways (in one or all of them simultaneously): 1) purchase of foreign currencies (US dollars, euro, GB pounds) and allocation to the Federal Treasury's accounts with the Bank of Russia which pays interest on them according to bank account agreement; 2) purchase of foreign currencies and financial assets denominated in Russian ruble and eligible foreign currencies (further – eligible financial assets) [2]. The allocation of National Wealth Fund's assets to deposits in Vnesheconombank is presented in Table 3.

The profitability of placing funds of the National Welfare Fund on an account in dollars last year amounted to 0.57% per annum (1.01% per annum since its inception), in euro - minus 0.76% per annum (1.28%); in pounds sterling - minus 0.16% per annum (2.91%).

The fund's return on deposits at Vnesheconombank in Russian rubles during this time was 6.44% per annum (6.64% from the commencement of deposit operations), on deposits in dollars - 0.25% (2.53%).

*Table 3. Allocation of National Wealth Fund's assets to deposits in Vnesheconombank as of November 1, 2018*

| ELIGIBLE INVESTMENTS OF NATIONAL WEALTH FUND'S ASSETS ALLOCATED TO DEPOSITS / PURPOSE OF ALLOCATION | MAXIMUM OVERALL AMOUNT OF ALLOCATION | ACTUAL AMOUNT OF ALLOCATION |
|---|---|---|
| Subordinated loans to Russian banks | 410.00 bln. rubles | 50.45 bln. rubles |
| - | - | including: |
| - | - | 21.27 bln. rubles |
| - | - | 29.18 bln. rubles |
| Not regulated | 175.00 bln. rubles | 50.00 bln. rubles |
| - | - | 15.98 bln. rubles |
| Loans to small and middle enterprises | 30.00 bln. rubles | 30.00 bln. rubles |
| Loans to joint-stock company 'DOM.RF Russia Housing and Urban Development Corporation' | 40.00 bln. rubles | 40.00 bln. rubles |
| Vnesheconombank's capital increase | 5.966 bln. US dollars (7% of overall volume of National Wealth Fund as of 6 September 2014) | (2.462 bln. US dollars) 138.83 bln. rubles |
| - | - | (3.504 bln. US dollars) 197.58 bln. rubles |
| Not regulated | No data | (0.288 bln. US dollars) 16.24 bln. rubles |
| Financing of Vnesheconombank's projects, implemented by organizations of real economy | 300.00 bln. rubles | 23.98 bln. rubles |
| - | - | 2.32 bln. rubles |
| - | - | 14.21 bln. rubles |
| **TOTAL:** | **955.00 bln. rubles** | **579.59 bln. rubles** |

*Source: [2].*

The yield of funds in securities of Russian issuers related to the implementation of self-financing infrastructure projects was as follows: in preferred shares of non-financial organizations - 0.49% per annum (0.60% per annum from the start of operations), in ruble bonds - 10.02% per annum (11.05% per annum from the start of operations), in dollar bonds - 3.65% per annum (3.36% from the start of operations), in preferred shares of banks - 4.86% per annum (2.1% from the beginning of operations).

According to the funds placed on deposits with VTB and Gazprombank to finance self-sustaining infrastructure projects, the yield was 8.27% per annum (10.39% per annum since the start of deposit operations).

From the moment the fund was created, the yield in the currency basket was 1.32% per annum, in rubles - 13.48%.

These data show different profitability from different kinds of investments.

## 5. Conclusion

The analysis of the main characteristics of the Government Pension Fund-Global of Norway showed, that the return on the portfolio in% per month is calculated based on the distribution of cash flows at the initial and final points in time.

Norway's experience in investing the assets of the sovereign fund may be useful in Russia. In particular, the fund of the Russia can also be invested in securities, taking into account several levels of diversification in order to accumulate higher profitability.